\documentclass[twocolumn,secnumarabic,amssymb, amsmath, nofootinbib,tightenlines,
nobibnotes, aps, prl]{revtex4}
\usepackage{amsmath}%
\usepackage{longtable}%
\usepackage{bm}%
\expandafter\ifx\csname package@font\endcsname\relax\else
 \expandafter\expandafter
 \expandafter\usepackage
 \expandafter\expandafter
 \expandafter{\csname package@font\endcsname}%
\fi

\usepackage{graphicx}
\usepackage{graphicx}

\begin{document}

\title{Lagrangian Structure Functions in Turbulence: Scaling Exponents and Universality}

\author{Victor Yakhot}
\email{vy@bu.edu}
\affiliation{Department of  Mechanical Engineering, \\
Boston University, Boston, MA 02215}
\date{today}%

\begin{abstract}
\noindent    In this paper, the approach for investigation of asymptotic ($Re\rightarrow \infty$) scaling exponents of Eulerian structure functions  (J. Schumacher et al,  New. J. of Physics {\bf 9},  89  (2007). ) is generalized to studies of Lagrangian structure functions in turbulence.  
The novel "bridging relation" based on   the derived expression for the fluctuating,  moment-order - dependent dissipation time $\tau_{\eta,n}$,   led to analytic expression for  scaling exponents 
($\kappa_{n}$ ) of the moments of Lagrangian velocity differences $S_{n,L}(\tau)=\overline{(u(t+\tau)-u(t))^{n}}\propto \tau^{\kappa_{n}}$ in a good agreement with experimental  and numerical data. 
\end{abstract}
\maketitle

\noindent  {\it Introduction.} %The almost hundred -years -long  effort directed to quantitative description  of geometry, dynamics and  engineering applications of turbulence,  led to various novel and important concepts, such as dynamic scaling and mode coupling, which are at the foundation of modern theory of phase transitions, renormalization group etc.  These methods inspired  formulation of turbulence models for momentum, heat and mass transfer (mixing) which at the present time evolved into one of the most important elements of the engineering design cycle.  Despite these successes,  many  important unsolved problems  remain.  
 A turbulent flow can be described  using Eulerian and Lagrangian approaches addressing  dynamics of  velocity field and evolution  of individual fluid particles, respectively. Therefore,  one can introduce  two kinds of structure functions, i.e. moments of velocity increments.     The properties of  Eulerian correlation functions  (ESF)  were first   theoretically  investigated by Kolmogorov in his celebrated K41 theory of turbulence  based on an exact in the inertial range relation for the third-order structure function  $S_{3}(r)=\overline{{\bf (u(x+r)-u(x))}\cdot \frac{{\bf r}}{r})^{3}}\equiv \overline{({\bf \delta_{r}u}\cdot  \frac{{\bf r}}{r})^{3}}\propto r$ [1].   The problem  of Lagrangian correlation functions,  for which not a single exact dynamic relation exists,  is not new and was originally  formulated within the framework of Kolmogorov theory of turbulence (for a detailed review, see Ref.[2]).  Due to technological limitations of the past,  experimental studies of strong turbulence were mainly devoted to   Eulerian structure functions (ESF)  $S_{n}=\overline{(u(x+r)-u(x))^{n}}\equiv\overline{(\delta_{r}u)^{n}}$  limited to the single- point measurements with subsequent application of Taylor hypothesis.   Here $u$  is a  component of velocity field parallel  to the displacement vector ${\bf r}$ chosen along the $x$-axis.  The experimental and numerical  studies   of Eulerian structure functions   revealed two distinct intervals:  while in  the analytic range $r\rightarrow 0$ (or  $r\ll \eta_{n}$),  the velocity field is differentiable  and $S_{n}(r)\propto r^{n}$,  at  the scales $L\gg r \gg  \eta_{n}$,  the ESFs  are given by algebraic  relations $S_{n}(r)\propto r^{\xi_{n}}$  with `anomalous' scaling  exponents $\xi_{n}$.  By this definition,  $\eta_{n}$'s are the cut-offs separating analytic and `rough'  ranges of ESFs.   (See Fig.1). 

\noindent The remarkable  breakthroughs in experimental particle tracking, led by the Bodenschatz group [3]-[7],   enabled  investigation  of Lagrangian structure functions  (LSF) $S_{n,L}=\overline{(u(t+\tau)-u(t))^{n}}\equiv\overline{(\delta_{\tau} u)^{n}}$ which are a crucial ingredient in understanding of  turbulent transport and mixing.  Due to analyticity of velocity field, as $\tau\rightarrow 0$,  $\delta_{\tau}u\approx a\tau$ where $a$ is  acceleration of a fluid particle and  $S_{nL}=\overline{a^{n}}\tau^{n}$. For  the time-increments $\tau\gg \tau_{\eta,n}$, the LSFs $S_{nL}\propto \tau^{\kappa_{n}}$ with the dissipation times $\tau_{\eta,n}$ separating analytic and rough intervals on a time-domain.  (See Fig.1). 
The recent  multifractal theory of LSFs [6] stressed the importance of the dissipation time $\tau_{\eta}$  which was treated in the spirit of Kolmogorov theory as a moment-number- independent quantity $\tau_{\eta}\propto T/\sqrt{Re}=const$ where $T\approx L/u_{rms}$ is a large-scale eddy turn-over time.  
To make a connection between ESFs  and  LSFs,  the authors of Ref. [6]  used  the  "bridging relation" (BR):
$r\approx \tau\delta_{\tau} u\equiv \tau (u(t+\tau)-u(t))$ with the time-increment $\tau$ approximately equal to the fluctuating  eddy turn-over time in {\it the inertial range}.   This expression, introduced on dimensional grounds  in Ref.[8],   can be understood as follows.  Consider a fluid particle at a time $t=t_{0}$ occupying position ${\bf X}_{0}$. Then, the particle displacement is: ${\bf R}={\bf X}(t_{0}+\tau)-{\bf X}_{0}=\int_{t_{0}}^{t_{0}+\tau}{\bf u}(\lambda)d\lambda$, so that 
${\bf R}={\bf u}(t_{0})\tau +\int_{t_{0}}^{t_{0}+\tau} ({\bf u}(\lambda)-{\bf u}(t_{0}))d\lambda$. 
We can see that if ${\bf u}(t)={\bf U}=const$,  the displacement ${\bf R}\neq 0$. Then, following [2] (page 342, where it is related to transition to a moving  frame of reference), we define the
regularized (not involving single-point- velocity)   quantity:  ${\bf r}={\bf R}-{\bf u}(t_{0})\tau$.   (It  is estimated in Ref.[2] (page 359) 
that  $r\approx  0$,  meaning that it tends to zero when either  $u\rightarrow U=const$ or  $\tau\rightarrow 0$.)  From the mean value theorem we also have $\int_{t_{0}}^{t_{0}+\tau} u(\lambda)d\lambda\approx u(t+\overline{\tau})\tau$, where $0<\overline{\tau}\leq \tau$. The  relation   $ r\approx \tau\delta_{\tau}u$ is obtained in the first approximation setting  $u(t+\tau)\approx u(t+\overline{\tau})$ which resembles the one used in construction  of Kraichnan's  Lagrangian History Direct Interaction Approximation [9].  Keeping in mind topological complexity of developed turbulence, we conclude that the mean value theorem,  leading to the "bridging relation" $r\approx \tau\delta_{\tau}u$ with the inertial range time-increments $\tau\gg \tau_{\eta}$,  cannot be accurate.
 Recently, the  BR   was analyzed using the exact relations  between Eulerian and Lagrangian structure functions by Kamps et al [10] who showed that,  combined with the multifractal formalism, it leads  to the  Lagrangian exponents  in a substantial disagreement  with experimental [3]-[7] and numerical [11] data.  Moreover,  it was pointed out  that  the theory of Ref. [6],   expressing   anomalies of  Lagrangian  exponents in terms of  anomalies of   Eulerian ones,    does not explain the "2D-paradox":   while the  two-dimensional Eulerian turbulence is not intermittent,  the Lagrangian one is.

\noindent In this paper, based on the ideas developed in Refs.[12]-[16],  we attack the problem differently. In the {\it analytic (dissipation)}  range   where   $\delta_{\tau}u\approx a\tau$,   the displacement  $r\approx \tau\delta_{\tau} u/2$.   Extrapolating this to the dissipation cut-off  $\eta$,  separating analytic and rough intervals  of the structure functions,  leads to the dissipation time $\tau_{\eta}$ :

\begin{equation}
\eta\approx \tau_{\eta}\delta_{\tau_{\eta}}u
\end{equation}

\noindent Both $\eta\approx \frac{\nu}{\delta_{\eta}u}$ and $\tau_{\eta}$ in formula (1)  are random functions investigated in great detail theoretically and numerically [13]-[16]. 
Unlike the BR defined for the large inertial range values of the time - increment $\tau \gg \tau_{\eta}$, the  expression (1), introduced as  an extrapolation of the  exact in the analytic interval relation,  is accurate   
for  short times $\tau\leq \tau_{\eta}$.    {\it Theoretical basis for this extrapolation is understood as follows. The structure functions, both Eulerian and Lagrangian,  can be formally represented as $\psi_{n}(x)\propto x^{\beta_{n}(x)}$  with the $x$-dependent exponents $\xi_{n}\leq \beta_{n}(x)\leq n$ (or $\kappa_{n}\leq \beta_{n}(x)\leq n$)  covering both analytic  ($\beta_{n}=n$) and inertial ( $\beta_{n}(x)=\xi_{n}$ or $\beta_{n}=\kappa_{n}$) ranges.  The smallest inertial range  exponents giving maximum values to the differences $n-\beta_{n}$, correspond to the strongest singularities of velocity field and 
defining the cut - offs by the matching relation  $  \overline{a^{n}}\tau_{\eta,n}=K_{n}\tau_{\eta,n}^{\kappa_{n}}$,  we account for the strongest, dominant,  singularities.  (See Ref.[16]). }

\begin{figure}[!h]
{\includegraphics[angle=0,scale=1.1]{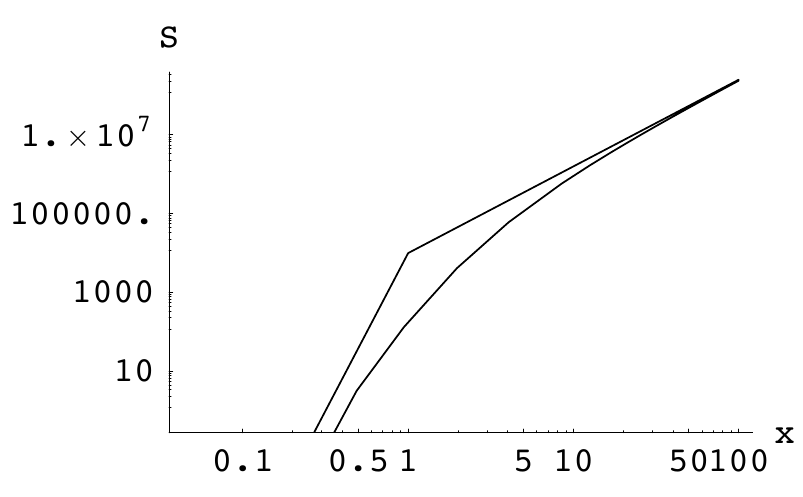}}
%{\includegraphics[angle=0,scale=0.5]{venturi_70.pdf}}
%\subfigure[]{\includegraphics[angle=0,scale=0.5]{w10_21s.eps}}
%\subfigure[]{\includegraphics[angle=0,scale=0.5]{w10_37s.eps}}
\caption{ Definition of the cut-off $\eta_{n}$ ( or $\tau_{\eta_{n}}$) for  a  structure function  $S_{n}(x)$  (or$(S_{n,L}(x)$).  As an example $S=S_{8}$ is chosen. The strait lines are $x^{8}$ (analytic range) and $x^{2.2}$ (inertial one). The point of their crossing is $x_{n}=\eta_{n}$.  Schematic.}
\label{fig} 
\end{figure}

\noindent The probability distribution functions $Q(\eta,Re)$, derived and investigated numerically in [14]-[15]   enable one to evaluate the moments of the dissipation scales:

\begin{equation}
\eta_{n}^{n}=\overline{\eta^{n}}=\int \eta^{n}Q(\eta,Re)d\eta \approx LRe^{\frac{n}{\xi_{n}-\xi_{n+1}-1}}
\end{equation}

\noindent Combining (1) and (2)  and assuming continuity of  $S_{nL}(\tau)$ on the dissipation time scale yields:

\begin{equation}
L^{n}Re^{\frac{n}{\xi_{n}-\xi_{n+1}-1}}=(u_{rms}T)^{n}(\frac{\tau_{\eta,n}}{T})^{\kappa_{n}+n}
\end{equation}

\noindent   Our goal now is to express the dissipation times $\tau_{\eta,n}\equiv \tau_{n}$ in terms of the Reynolds number and,   comparing the result with formula (2),  obtain expressions for Lagrangian exponents $\kappa_{n}$.

%In this paper we  argue  that to quantitatively describe  LSFs one has to accurately account for the strong,  intermittency-induced,  fluctuations of the dissipation time. Due to intermittency,  the velocity fluctuation on a spatial scale $\eta_{n}$ is  dissipated on its  `own'  $n$-dependent dissipation time -scale $\tau_{\eta,n}\equiv \tau_{n}$.  {\it Using this concept, we are able to derive a set of novel  "bridging relations" corresponding to the uv   time - increments $ \tau_{n}$ which are small, thus leading to a  much more accurate application of the  mean-value theorem discussed above.  }
%This new concept leads to the closed expressions   for the scaling exponents of Lagrangian structure functions in a quantitaive agreement with experimental data.  The proposed theory, which  is valid for  3D turbulence only,  sheds  light on  the 2D paradox discussed in Ref. [10]. 

\noindent {\it Eulerian structure functions.}  Since  energy dissipation in the inertial range dynamics   ( inverse-energy-cascade ) of  2D turbulence is irrelevant,  the relation  (2), obtained by balancing  viscous dissipation and inertial-range contributions to  the exact  equations for  the Eulerian structure functions (see Refs. [14]-[16]) is valid for three-dimensional flows only. 
%Therefore, the expression (2) and the theory developed below are applicable to 3D turbulence only. 
If, in accord with Kolmogorov theory, we  assume $\xi_{n}=n/3$, this formula  gives the well-known  $n$-independent relation $\eta_{n}\approx L Re^{-\frac{3}{4}}$.  However, due to intermittency, 
the function $\xi_{n}$ is a convex function of the moment order  $n$ and for a  fixed Reynolds number $Re=const$,  the dissipation scales $\eta_{n}$ decrease with $n$.  (This result has been  numerically tested  in Ref.[16]).
 Since the inertial range is compressed to the interval between integral and dissipation scales ($\eta_{n}\ll r\ll L/10$), experimental determination of exponents $\xi_{n}$ is very difficult and at the present time 
only the exponents $\xi_{n}$ with $n\leq 8$ have been  accurately established by direct investigation of inertial range dynamics.  In accord with the recently developed theory of small-scale intermittency,  the dissipation scale is defined by a dynamic Reynolds number $Re_{\eta}=\frac{\eta\delta_{\eta}u}{\nu}\approx 1$ [15].   First, we see that the dissipation scale is not a constant as in K41, but a fluctuating property of a flow. Then, it is easy to show [14[-[16] that $
\frac{\partial u}{\partial x}=\frac{\delta_{\eta}u}{\eta}\approx \frac{(\delta_{\eta} u)^{2}}{\nu}$.  Since $\eta$ is the scale separating analytic and rough scale- intervals of the velocity field, it has been shown that:

\begin{equation}
\overline{(\frac{\partial u}{\partial x})^{n}}\approx S_{2n}(\eta_{2n})/\nu^{n}\propto  \eta_{2n}^{\xi_{2n}}/\nu^{n}\propto Re^{\rho_{n}}
\end{equation} 

\noindent Using (2) we derive readily:

\begin{equation}
\rho_{n}=n+\frac{\xi_{2n}}{\xi_{2n}-\xi_{2n+1}-1};~d_{n}=n+\frac{\xi_{4n}}{\xi_{4n}-\xi_{4n+1}-1}
\end{equation}

\noindent  where $d_{n}$ are the exponents of  the moments of the dissipation rate $\overline{{\cal E}^{n}}\approx \frac{\overline{ (\delta_{\eta}u)^{4}}}{\nu}=S_{4n}(\eta_{4n})/\nu\propto Re^{d_{n}}$. 
According to this calculus, the moments of Lagrangian acceleration 

\begin{equation}
\overline{a^{n}}=\overline{(\frac{(\delta_{\eta}u)^{3}}{\nu})^{n}}\propto Re^{\frac{\xi_{3n}}{\xi_{3n}-\xi_{3n+1}-1}+n}\equiv Re^{\alpha_{n}}
\end{equation}

\noindent The  quantitative results  of Ref. [16], obtained with the help of a particular parametrization  (Ref.[13]-[15])
\begin{equation}
\xi_{n}=0.383 n/(1+n/20)
\end{equation}

%\section{Moments of velocity derivatives.}
%\setcounter{equation}{1}
%\setcounter{equation}{0}
\noindent  The predicted in Refs. [13]-[16] relations (4),(5)  have been  confirmed in  the  most detailed numerical simulation of Ref.[16] and [17]. {\it The expression (7),  calibrated to give $\xi_{3}=1$,  is the result of a  theory valid for the even - order moments only.  We do not have any reason to believe that $\overline{|\delta_{r}u|^{3}}\propto r^{\xi_{abs}}$  with $\xi_{abs}=1$.  Thus, the accuracy of the relation (7) must be of the order  $|\xi_{3}-\xi_{3,abs}|$,  which is numerically small. This drawback is common to  all exisitng  models leading to  expression for $\xi_{n}$. }
% high resolution  ($N=512^{3}-2048^{3}$) 
%numerical simulations  of moments of velocity derivatives $\overline{(\frac{\partial u}{\partial x})^{n}}$ and dissipation rate  $\overline{{\cal E}^{n}}$ in  isotropic and homogeneous turbulence in
 %the relatively ``low Reynolds number'' range $R_{\lambda}=10-107$  by Schumacher et al and $R_{\lambda}\approx 107-140$ by  Donzis et al [11].  In all cases, to precisely  evaluate velocity derivatives $\partial_{x}u$,   a special care was taken to accurately resolve the analytic ranges of  corresponding structure functions $S_{n,0}\propto r^{n}$.  The results are are shown on Fig.1 where the numerical data on the moments of derivative (Fig. 1a) and dissipation rate  (Fig. 1b) are compared with predictions (3),(4),(5) .   As we see,  the  well-defined algebraic ranges with exponents calculated and measured with unprecedented precision,  can be detected in these low Reynolds number, inertial range - lacking,  flows.   (The  high-order, high Reynolds number ($n=4$, $Re_{\lambda}=140$)   moments  shown on the  Fig. 1 are slightly underesolved.)

\noindent {\it Lagrangian structure functions.}  The classical treatment of the problem can be formulated as follows [2].  
As  $\tau\rightarrow 0$,   the velocity field is differentiable  and using (6):

\begin{equation}
S_{2n,L}(\tau)= \overline{(\delta_{\tau}u)^{n}}=\overline{a^{2n}}\tau^{2n}\propto Re^{\alpha_{2n}}\tau^{2n}
\end{equation}
 
\noindent  The  limit $\tau\rightarrow 0$,   means that there exist a time-scale $\tau_{\eta,2n}$ such  that  the relation (8) is valid in the interval $\tau\ll \tau_{\eta,2n}$.  In the "inertial range $\tau_{\eta,2n}\ll \tau \ll T=\frac{1}{u_{rms}^{2}}\int_{t_{0}}^{t_{0}+\tau}\overline{u(0)u(t) }dt$, the velocity field is not differentiable and 

\begin{equation}
S_{2n,L}(\tau)\propto \tau^{\kappa_{2n}}
\end{equation}

%\begin{figure}
%\hspace{4cm}
%\subfigure{\includegraphics[height=4cm]{MAT_SN4B.eps}}
%\includegraphics[scale=0.65]{MAT_PDFB.pdf}
%\includegraphics[scale=0.4]{mat_pdfd_naphys.pdf}
%\subfigure[]{\includegraphics[height=5cm]{ESSs162_gr3.pdf}}   \subfigure[]{\includegraphics[height=5cm]{ESSs16_gr4.pdf}} 
%\centerline{\includegraphics[angle=0,scale=0.7,draft=false]{ESSFig1_gr6.pdf}}
%\centerline{\includegraphics[angle=0,scale=0.7,draft=false]{ESSFig1_gr5.pdf}}
%\centerline{\includegraphics[angle=0,scale=0.7,draft=false]{ESSFig1_gr3.pdf}}
%\caption{ Moments of velocity derivatives~ (a) ~and dissipation rate ~ (b)~ vs, large-scale Reynolds number.  (Squares~-~the data from Schumacher et al; $*$~-~simulations of Donzis et al. of Ref.[11])}
%\label{fig1}
%\end{figure} 

%\noindent The excellent agreement between theory and numerics of Ref.[11] demonstrate a relatively good understanding of Eulerian structure functions. 

\noindent   In what follows we define n$^{th}$ dissipation time $\tau_{2n}$ by the matching relation:
  
\begin{equation}
S_{2n,L}(\tau_{2n})=\overline{a^{2n}}\tau_{2n}^{2n}=({\cal E }\tau_{2n})^{n}(\frac{\tau_{2n}}{T})^{\kappa_{2n}-n}
\end{equation}

\noindent  {\it According to  K41}, neither integral scale nor viscosity  influence the dynamics of inertial range and,  on dimensional grounds $\kappa_{2n}=n$ and 
$S_{2n,L}=K_{2n}({\cal E}\tau)^{n}$ with $K_{n}=const.$
Let us examine consistency of this  result  with some other predictions of the K41. It is clear that:
$S_{2,L}=\int_{0}^{\tau}\int_{0}^{\tau}\overline{a(\lambda_{1})a(\lambda_{2})} d\lambda_{1} d\lambda_{2}$. By the time-homogeneity $\overline{a(\lambda_{1})a(\lambda_{2})}=A(\lambda_{1}-\lambda_{2})\equiv A(s)$ and

\begin{equation}
S_{2,L}=\int_{0}^{\tau}(\tau-s)A(s)ds
\end{equation}

\noindent According to K41 (see Ref.[2]),  in the interval $0<\tau <\tau_{\eta,2}$,  $A(\tau)\approx\overline{a^{2}}$  and 
for $\tau_{\eta,2}<\tau \ll T$,  $A(\tau)\approx K_{2}{\cal E}/\tau$.  Substituting this into (11) gives the inertial range expression  ($\tau\geq \tau_{\eta,2}$)

\begin{equation}
S_{2,L}(\tau)\approx \frac{\overline{a^{2}}\tau_{2}^{2}}{2}+K_{2}{\cal E}(\tau\ln\frac{\tau}{\tau_{2}}-\tau+\tau_{2})
\end{equation}

\noindent  We can notice a slight  inconsistency of K41:  the expressions (12) and the K41 result $S_{2,L}\propto \tau$  differ by a large logarithmic factor which is impossible to derive from  dimensional considerations. 
Dimensional considerations  [2] also   give $\overline{a^{2}}\approx {\cal E}^{\frac{3}{2}}/\sqrt{\nu}\propto \frac{u_{rms}^{4}}{L^{2}}\sqrt{Re}$  
and up to logarithmic correction, the second contribution to the right side of (12) is $O(1)$.  Demanding continuity of the structure function $S_{n,L}$ in the limit $Re\rightarrow \infty$, we obtain the familiar K41  estimate for the relaxation time $\tau_{2}\approx \frac{u_{rms}}{a_{rms}}\propto Re^{-\frac{1}{2}}$.  
{\it Accounting for intermittency.}   Since,  $T\approx L/u_{rms}$ , then, as  follows from (10), on the dissipation time-scale:

\begin{equation}
s_{2n}(\tau_{2n})=\frac{S_{2n,L}(\tau_{2n})}{u_{rms}^{2n}}=Re^{\alpha_{2n}}(\frac{\tau_{2n}}{T})^{2n}=K_{2n}(\frac{\tau_{2n}}{T})^{\kappa_{2n}}
\end{equation}

\noindent and for $\tau\ll \tau_{2n}$:

\begin{equation}
S_{2n,L}\propto Re^{\alpha_{2n}}\tau^{2n}; \hspace{0.5cm} \frac{\tau_{2n}}{T}\approx  Re^{\frac{\alpha_{2n}}{\kappa_{2n}-2n}} \equiv Re^{-\gamma_{2n}}
\end{equation}

\noindent 
Now, we derive the expression for $\kappa_{n}$.  Combining (2),(3) with (14) gives:

\begin{eqnarray}
L^{n}Re^{\frac{n}{\xi_{n}-\xi_{n+1}-1}}=(u_{rms}T)^{n}(\frac{\tau_{n}}{T})^{\kappa_{n}+n}=\nonumber \\
(Tu_{rms})^{n}Re^{\frac{\alpha_{n}(\kappa_{n}+n)}{\kappa_{n}-n}}
\end{eqnarray}

\noindent  and 

\begin{equation}
\frac{n}{\xi_{n}-\xi_{n+1}-1}=\frac{\alpha_{n}(\kappa_{n}+n)}{\kappa_{n}-n}
\end{equation}

\vspace{0.5cm}

% \begin{indented}
% \lineup
% \item[]\begin{tabular}
% \br
%\begin{array}{clcr}s

%n & \xi_{n} & \alpha_{n} & \kappa_{n} \\
%2 & 0.7 & 0.55  & 0.99  \cr
%4  &1.27 &  1.5 & 1.35 \cr
%6 & 1.77 & 2.6 & 1.73 \cr
%8 & 2.2 & 3.9 & 2.1 \cr
%10 & 2.55 & 5.8 & 2.35 \cr 
%\end{array}
%\br
%\end{tabular}
%\end{indented}
%\end{table}
%"""""""""""""""""""""""""""""""""
%\begin{indented}
%\lineup
%\item[]\begin{tabular}{@{}*{6}{l}}
%\br
%\begin{array}
% n & \xi_{n}& \alpha_{n}& \kappa_{n}^{th}&\kappa_{n}^{exp}&\gamma_{n}\cr
%\mr
%$2$ & 0.7& 0.55  & 0.99 &1. & 0.545\cr
%$4$ & 1.27 & 1.5 & 1.35 &1.3-1.6 [2], [5], [11] & 0.566 \cr
%$6$ & 1.77 & 2.6 & 1.73 &1.6-1.8 [2],[5],[11] & 0.61\cr
%$8$ & 2.2 & 3.9 & 2.1& na &  0.66\cr
%$10$ & 2.55 & 5.8  & 2.35& na &  0.76\cr\mr
%\end{array}
%\br
%\end{tabular}
%\end{indented}
%\end{table}

\renewcommand{\arraystretch}{1.8}
\begin{table}
\begin{center}
\begin{tabular}{lcccccccc}
\hline\hline
n& $\xi_{n}$ & $\alpha_{n}$ & $\kappa_{n}^{th}$ & $\kappa_{n}^{exp}$ & $\gamma_{n}$ \\
\hline
2   & 0.7  & 0.55  & 0.95                     & 0.9-1. [4]'[10[,[11]  & 0.545     \\
4   & 1.27 & 1.5 & 1.44                    & 1.3-1.6 [2],[4],[5],[11]   & 0.566    \\
6   & 1.77 & 2.6 & 1.74                     & 1.6-1.8 [2],[4],[5],[11]    & 0.61  \\
8   & 2.2 & 3.9 & 1.92                     & 1.9,[4]    &0.66   \\
10 &2.55 &   5.8       & 2.05& 2.1, [4]  & 0.76\\
\hline\hline
\end{tabular}
 
\end{center}
\label{tab1}
\end{table}

\begin{figure}[!h]
{\includegraphics[angle=0,scale=1.15]{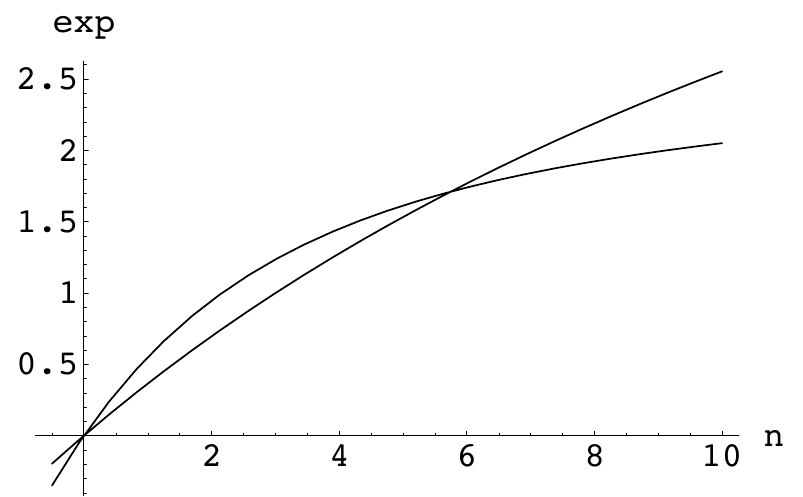}}
{\includegraphics[angle=0,scale=0.5]{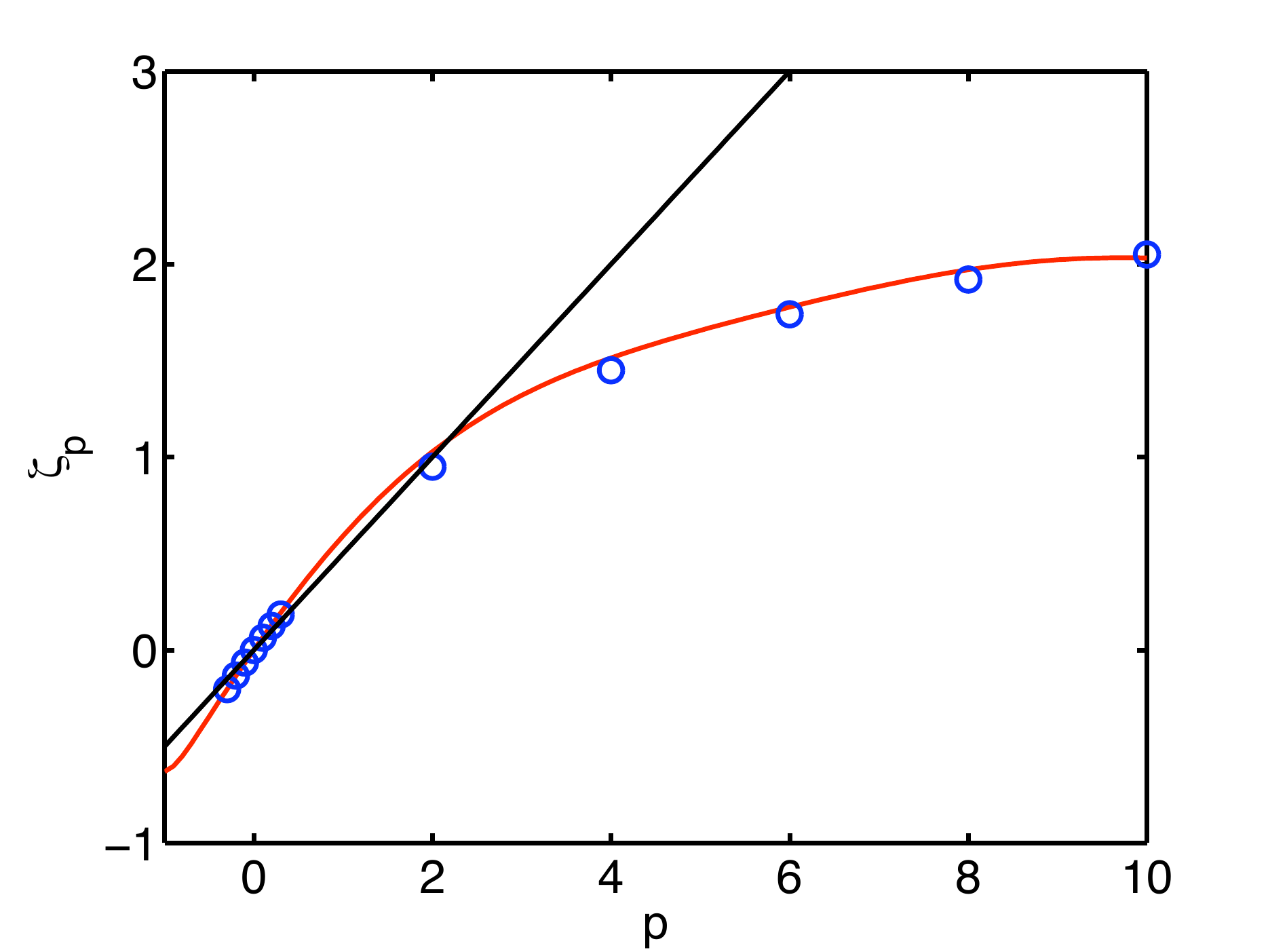}}
%\subfigure[]{\includegraphics[angle=0,scale=0.5]{w10_21s.eps}}
%\subfigure[]{\includegraphics[angle=0,scale=0.5]{w10_37s.eps}}
\caption{ Top curve:  Lagrangian and Eulerian exponents $\kappa_{n}$ and $\xi_{n}$  vs. $n$.  Lagrangian ones are "more intermittent".  Bottom: comparison of predicted (formula (14)) and experimentally observed exponents  
$\kappa_{p}\equiv \zeta_{p}$ [4]. Black line is exponents based on  K41.}
\label{fig} 
\end{figure}

\noindent With the parametrization  (7) (or any other, for that matter) 
one is able to calculate the exponents $\kappa_{n}$ of Lagrangian structure functions.  The theoretical predictions ($\kappa_{n}^{th})$ are favorably compared with experimental and numerical ($\kappa_{n}^{exp})$ results  in the Table. 
%together with the  calculated exponents $\gamma_{n}$ of the dissipation times given by (14). The predicted 
%exponents $\gamma_{n}$ of the dissipation time, presented in the last column of the Table,  show 
%$\tau_{\eta,n}<\tau_{K}\propto T/\sqrt{Re}$.  Due to the  difficulties with particle tracking, investigation of LSF is very complicated and the measured scaling exponents $\kappa_{n}$ presented in the Table reflect large experimental error bars. For example, according to Ref.[11], the experimental  $\kappa_{4}=1.58\pm 0.06$  and $\kappa_{6}=1.9\pm 0.2$ [5],  and $\kappa_{4}=1.47\pm 0.38$  and $\kappa_{6}=1.66\pm 0.53$ [4] while numerical results are 
%$\kappa_{4}=1.46\pm0.06$, $\kappa_{6}=1.67\pm 0.19$ [11] and $\kappa_{4}=1.5\pm0.09$,$\kappa_{6}=1.67\pm0.2$.  

The  moment-number dependence of   Lagrangian and Eulerian exponents is presented on Fig.2 (top curve) together with comparison of theoretical predictions with experimental data of Ref. [4].

% For example,  for $Re\approx 10^{4}$, $\frac{\tau_{6}}{\tau_{K}}\approx 1/2.8$ while for $Re\approx 10^{5}$ this ratio is equal to $\approx 1/3.5$.  For $n=8$ and $n=10$, the ratios become even smaller.
%Our  predictions can be compared  with the experimental data on $\kappa_{n}$ available for  $2\leq n\leq 6$. 
% We see that in the limit $\tau\rightarrow 0$, the Lagrangian structure functions scale with 
%with the moments of acceleration which is a strongly intermittent property defined by the exponents  $\alpha_{n}$.  According to the theory presented in this paper,  $\alpha_{2}=0.55$, $\alpha_{4}=1.5$, $\alpha_{6}=2.6$ which differ from the  corresponding K41 values $0.5;~1.0~1.5$, respectively. 

\noindent The non-trivial Reynolds number dependence of the relaxation times $\tau_{\eta,n}$,  given by expression (14),  is also of importance. Plotting the  experimental data  on  $S_{n,L}(\tau)$  in coordinates $\tau/\tau_{K}\approx T/\sqrt{Re}$  instead of $\tau/\tau_{n}$, 
the authors of Ref. [7]  failed to  collapse the experimental graphs for $S_{nL}(\tau,Re)$, which indicated a dynamic inconsistency  of Kolmogoorv's dissipation time.    Similar phenomenon was described in  Schumacher et. al. (Ref.[16])  dealing with the moments of increments Eulerian velocity field. 
%The experimental  values for exponents $\kappa_{2}=1$, $\kappa_{4}\approx 1.4-1.5$ and \kappa_{6}\approx 1.7-1.8$ [2]-[5] are in a reasonable agreement with the ones predicted in the Table. 
%A  somewhat larger magnitudes  of exponents ($\kappa_{4}\approx 1.8; \kappa_{6}\approx 2.2$) resulting  from numerical simulations of Ref. [14] deserve  a special discussion.  Two flows ($Re=\frac{u_{rms} L}{\nu}\approx 10^{4}$)
%were generated in a periodic box and the probability density of normalized acceleration was plotted in the interval 
%$0\leq |\frac{a}{a_{rms}}|\leq 30}$. Based on the data presented in the Table, for this Reynolds number 
%$\frac{\overline{a^{4}}}{a_{rms}^{4}}\propto Re^{0.4}\approx 40$ (basically identical  to the data of Ref.  [14]) and $\frac{\overline{a^{6}}}{a_{rms}^{6}}\propto Re^{0.95}\approx 6000$. In Ref. [14] the exponents  $\kappa_{n}$ 
%were established using two different approaches:  the ESS method $S_{2n,L}=S_{2,L}^{\kappa_{2n}(ess)}$  and directly from the data. One can see from  the inserts  to Fig. 4 of Ref.[14]  that in the interval $\tau/\tau_{K}>1$ the exponents $\kappa_{6}$,  found from two different methods,  are very different with $\kappa_{6}(ess)>2$ and $\kappa_{6}\approx 1.8$ close to experimental and predicted numbers of the Table. This may point to  poor performance of the ESS methodology also reported by Schumacher et al in Ref.[11] .  

{\it Conclusions.}  In this paper we used the theoretically predicted and experimentally verified fact [16] that the inertial-range asymptotic exponents of Eulerian structure functions are closely related to the anomalous Reynolds -number- scaling  of the moments of {\it velocity derivatives}  defined on the fluctuating ultra-violet cut-offs given by expression (2).  Following [15],[16],  we  introduced the  moment-number-dependent dissipation time which,  combined with the new "bridging relation" $\eta\approx \tau_{\eta}\delta_{\tau_{\eta}}u$,  enabled  us to evaluate the scaling exponents of Lagrangian structure functions in a good agreement with experimental and numerical data. Unlike the previously used BR,  defined   for the inertial range increments $\tau\gg \tau_{\eta}$,  this relation, which is an extrapolation of an exact in the analytic range dependence $r=\tau \delta_{\tau}u/2$,   is much more accurate. 
%The theory presented here is applicable exclusively to three-dimensional turbulence where, due to dissipation anomaly [14], [16],  fluctuations of the energy dissipation play a  crucial role in the inertial range dynamics. In the two-dimensional turbulence on an infinite domain, the energy dissipation is irrelevant,  which explains the lack of anomalous scaling of Eulerian structure functions.  However, the small-scale low-energy 2D vortices, often appearing in the enstrophy-cascade  interval,  can serve as traps for the fluid particles,  thus leading strong Lagrangian intermittency. \\ 

I am grateful to Rainer Grauer for bringing my attention to the "2D-paradox" and for many illuminating discussions. 
Working on this paper I, as always,  benefited  from the evergreen volumes by A.S. Monin and A.M. Yaglom.  Sadly, this time I was unable  to discuss all this with Akiva Moiseevich whom we lost in the fall of 2007.  Most interesting and stimulating discussions with E. Bodenschatz,  L. Biferale, Haitao Xu, K.R. Sreenivasan, and J. Schumacher   are  gratefully acknowledged. 
 
 \section*{References}
%\begin{thebibliography}{100}
\noindent 1.~ A.N. Kolmogorov,    Dokl. Akad. Nauk SSSR  {\bf 30}, 299 (1941).\\
2.~A.S. Monin and A.M. Yaglom,   {\em Statistical Fluid
Mechanics,} vol.\ 2 1975, MIT Press, Cambridge, MA \\
3.~A. LaPorta et. al., Nature, London {\bf 409}, 1017 (2001).\\
4.~H. Xu, N. Ouelette and E. Bodenshcatz, Phys.Rev.Lett., {\bf 96}, 024503 (2006); 
5.~M. Mordant, P. Metz, O. Michel and J.-F.  Pinton, Phys.Rev.Lett. {\bf 87}, 214501 (2001).\\
6.~A. Arneodo et. al., Phys.Rev.Lett. {\bf 100}, 254504 (2008.)\\
7.~L. Biferale, E. Bodenschatz, M. Cencini, A.S. Lanotte, N.T. Oulette, F. Toschi, H. Xu,  Phys. Fluids {\bf 20}, 065103 (2008)\\
8.~M.S. Borgas, Phil. Trans. R. Soc. {A 342}, 379 (1993);  G. Boffetta, F. De Lillo, and S. Musacchio, Phys. Rev. {\bf E 66}, 066307 (2002);\\
9.~R. H. Kraichnan, Phys. Fluids {\bf 8}, 575 (1965).\\
10.~O.Kamps, R. Friedrich and R. Grauer,  arXiv:0809.4339[physics.flu-dyn].\\
11.~H. Homann, R. Grauer,  A. Busse and W.C. Muller, J.Plasma Phys.  {\bf 73}, 821 (2007).\\
12.~G. Paladin  and A. Vulpiani,  `Anomalous scaling laws in multifractal objects', Phys.Rep. {\bf 156}, 147-225 (1987) .\\
Phys. Rep. {\bf 35} 1971-1973 (1987); ~M. Nelkin, Phys. Rev. A {\bf 42} 7226 (1992); ~ U. Frisch and M. Vergassola, Europhys. Lett. {\bf  14} 439 (1991).\\
13.~V. Yakhot,   `Pressure-velocity correlations and anomalous exponents of structure functions in turbulence',  J. Fluid Mech.,  {\bf 495}, 135 (2003).\\ 
14.~V. Yakhot  and K.R. Sreenivasan,  `Toward dynamical theory of multifractals in turbulence',  Physica A {\bf 343}, 147-155. Physica A {\bf 343} 147 (2004);~ V. Yakhot, Physica D
{\bf 215} 166-174 (2006).\\
15.~V. Yakhot, J. Fluid. Mech. {\bf 606},  325 (2008).\\
16.~J. Schumacher, K.R. Sreenivasan and V. Yakhot, New. J. of Physics {\bf 9}, 89 (2007).\\
17.~D.A. Donzis, P.K. Yeung and K.R. Sreenivasan,   Phys.\ Fluids, {\bf 20}, 
045108, (2008).\\

%14.~L. Biferale, G. Boffetta, A. Celani, A. Lanotte and F. Toschi, arXiv: nlin/0402032v (2008);  
%  L. Biferale et al., Phys. Fluids 17, 021701 (2005).\\
\end{document}